\documentclass[12pt]{article}

\def\gev2{\mbox{GeV$^2$}}

\newcommand{\be}{\begin{equation}}
\newcommand{\ee}{\end{equation}}
\newcommand{\bq}{\begin{eqnarray}}
\newcommand{\eq}{\end{eqnarray}}
\newcommand{\pup}{p^\uparrow}

\newcommand{\bfk}{\mbox{\boldmath $k$}}
\newcommand{\bfy}{\mbox{\boldmath $y$}}
\newcommand{\bfp}{\mbox{\boldmath $p$}}
\newcommand{\bfx}{\mbox{\boldmath $x$}}
\newcommand{\bfr}{\mbox{\boldmath $r$}}
\newcommand{\bfpi}{\mbox{\boldmath $\pi$}}
\newcommand{\bftau}{\mbox{\boldmath $\tau$}}
\newcommand{\bfG}{\mbox{\boldmath $G$}}
\newcommand{\bfI}{\mbox{\boldmath $I$}}
\newcommand{\bfJ}{\mbox{\boldmath $J$}}
\newcommand{\bfS}{\mbox{\boldmath $S$}}
\newcommand{\bfP}{\mbox{\boldmath $P$}}

\begin{document}

\pagestyle{empty}
\baselineskip 24pt 


\vspace{1cm}

\begin{center} 
{\bf \Large 
Non-Standard Time Reversal  \\ and
Transverse Single-Spin Asymmetries}

\end{center}

\vspace{1cm}

\baselineskip 16pt 

\begin{center}

{\large M. Anselmino$^a$, V. Barone$^{b,a}$, 
A. Drago$^c$ and F. Murgia$^d$ \\}

\vspace{0.5cm}

$^a$Dipartimento di Fisica Teorica, Universit{\`a} 
di Torino \\
and INFN, Sezione di Torino, 10125 Torino, Italy \\
$^b$Di.S.T.A., Universit{\`a} del Piemonte Orientale 
``A.~Avogadro''\\
and INFN, Gruppo Coll. di Alessandria, 15100 Alessandria, Italy \\
$^c$Dipartimento di Fisica, Universit{\`a} di Ferrara \\
and INFN, Sezione di Ferrara, 44100 Ferrara, Italy \\
$^d$Dipartimento di Fisica, Universit{\`a} 
di Cagliari \\
and INFN, Sezione di Cagliari, 09042 Monserrato (CA), 
Italy

\end{center}

\baselineskip 16pt

\vspace{1cm}

\begin{center}
{\bf \large Abstract}
\end{center}

\noindent
A system of quarks interacting with chiral fields is shown to provide 
a physical realization of a ``non-standard'' time reversal for  
particle multiplets which mixes the multiplet components. We argue 
that, if the internal structure of the nucleon is governed by a chiral 
dynamics, the so-called $T$-odd quark distribution functions
are not forbidden by time-reversal invariance and hence might be non 
vanishing. This agrees with some other recent results. From a 
phenomenological point of view, this would give rise to single-spin 
asymmetries in inclusive processes involving a transversely
polarized nucleon: in particular, in pion lepto- and hadro-production 
and in Drell-Yan processes.

\newpage

\baselineskip 16pt
\pagestyle{plain}
 
 
\noindent
{\bf 1.} Time-Reversal (TR) invariance is a powerful constraint on many 
physical processes. In the context of hadronic physics, for instance, 
it limits the admissible forms of structure functions, quark distributions, 
form factors, decay observables, {\it etc.} In this paper we shall present a 
physical realization of a ``non-standard'' time reversal for particle 
multiplets proposed by Weinberg \cite{weinberg} and characterized 
by a mixing of the multiplet components. The physical system that will 
be considered is an isospin doublet of quarks interacting with 
time-independent chiral fields ($\pi$'s and $\sigma$'s). We shall show 
that this system of quarks and mesons is indeed invariant under the 
non-standard TR. As an interesting consequence, if the nucleon has a 
chiral dynamics and time-reversal invariance is implemented according to 
Weinberg's inversion operator for quark multiplets, some quark distribution 
functions that might appear to be time-reversal odd, do not actually conflict 
with TR invariance, and might therefore be non vanishing (a preliminary 
account of the ideas presented here was given in \cite{noi}).
Our conclusions agree with a recent paper by Collins \cite{collins2}, who 
also finds that time-reversal invariance -- when taking into account the 
path-ordered exponentiation of the gluon field in the operator definition 
of parton densities -- cannot forbid new kinds of spin and $\bfk_\perp$ 
dependent distribution functions. This important result opens the way to 
a rich and interesting phenomenology of transverse single-spin asymmetries.    

\vspace{0.5cm} 

\noindent
{\bf 2.} Acting on a momentum and spin eigenstate 
$\vert \bfp, j_3 \rangle$, the TR operator $T$ yields
\be
T \, \vert \bfp,  j_3 \rangle = (-1)^{j - j_3} \, 
\vert - \bfp, - j_3 \rangle \,, 
\label{TR1}
\ee
where $j$ is the particle's spin, $j_3$ its third component, and an 
irrelevant phase has been omitted. Recall also that $T$ maps ``in'' 
states into ``out'' states: $T : \vert {\rm in} \rangle 
\to \vert {\rm out} \rangle$.
Consider now a multiplet of particles labeled by some internal quantum 
number $a$. In the standard realization of TR, the $T$ operator is  
taken to be diagonal in $a$: 
\be
T \, \vert \bfp,  j_3, a \rangle = (-1)^{j - j_3} \, 
\vert - \bfp, - j_3, a \rangle \,.  
\label{TR1b}
\ee

In his Quantum Field Theory book \cite{weinberg} Weinberg 
has considered a more general possibility, namely that $T$ may mix the 
multiplet components (an idea originally due to Wigner \cite{wigner}). 
Thus a non-diagonal finite matrix ${\mathcal T}_{ab}$ appears in 
(\ref{TR1b}), which becomes
\be
T \, \vert \bfp, j_3, a \rangle = 
(-1)^{j - j_3} 
\sum_b {\mathcal T}_{ab} \, \vert - \bfp, - j_3, b \rangle \,.  
\label{TR5}
\ee
Since $T$ is antiunitary, ${\mathcal T}$ must be unitary. Weinberg 
has proven that the matrix ${\mathcal T}$ can be made block-diagonal
by a unitary transformation, with the blocks being either simple phases, 
or at most $2 \times 2$ matrices of the form
\be
\left(
\begin{array}{cc}
 0 & {\rm e}^{i\phi/2}  \\ 
 {\rm e}^{-i\phi/2} & 0  
\end{array}
\right) \,, 
\label{TR6}
\ee
where $\phi$ is a  real number.

Weinberg's ``non-standard'' time reversal may indeed be realized in quark 
phy\-sics, as we are now going to illustrate (hereafter we refer to 
(\ref{TR1b}) and (\ref{TR5}) as to the ``standard'' and ``non-standard'' 
TR, respectively).

Let us consider a $SU(2)$ chiral lagrangian describing the interaction 
of a quark field $\psi$ with two chiral fields $\sigma$ and $\bfpi$ 
\bq
{\mathcal L} &=& i\,\bar \psi \gamma^{\mu}\partial_{\mu} \psi
  - g \, \bar \psi \left( \sigma
  + i \, \gamma_5 \, \bftau \cdot \bfpi \right) \psi \nonumber \\
&+& \frac{1}{2} {\left( \partial_\mu \sigma \right)}^2
       + \frac{1}{2} {\left(\partial_\mu \bfpi \right)}^2
         -U\left(\sigma ,\bfpi\right)   \, ,
\label{TR7}
\eq
where $U(\sigma ,\bfpi)$ is a mexican-hat potential. Lagrangians of the 
type (\ref{TR7}) are widely used to describe the structure of hadrons
\cite{ripka}. For simplicity, we shall assume that the chiral fields 
represent a time-independent background (this is a common assumption   
in many chiral models of the nucleon, see {\it e.g.}, \cite{broniowski}).
The field equations obtained from the lagrangian (\ref{TR7}), in the mean 
field approximation, are \cite{broniowski}
\bq
\left [ i \, \gamma^{\mu} \partial_{\mu} 
- g \, (\sigma + i \, \gamma_5 \bftau \cdot 
\bfpi) \right ] \psi(x)  &=& 0 \, , 
\label{fieq1} \\
- \nabla^2 \bfpi + i \, N_c \, g \, \bar \psi(x) \gamma_5 
\bftau \psi(x) + \frac{\partial U}{\partial \bfpi} 
&=& 0 \,, 
\label{fieq2} \\
- \nabla^2 \sigma + N_c \, g \, \bar \psi(x)  
 \psi(x) + \frac{\partial U}{\partial \sigma} 
&=& 0 \,.
\label{fieq3}
\eq

Under TR the Dirac equation becomes (the superscript ``t'' denotes 
transposed)
\be
\left [ i \, \gamma^{\mu} \partial_{\mu} 
- g \, (\sigma - i \, \gamma_5 \bftau^{\rm t} \cdot 
\bfpi') \right ]  \gamma_5 \, {\mathcal C} \, \psi^*(\tilde x)  = 0 \, , 
\label{fieq4}
\ee
where $\tilde x=(-x_0, \bfx)$, ${\mathcal C} = 
i \gamma_2 \gamma_0$ and $\bfpi'$ is the time-reversed pionic field. 
Were pions absent, the time-reversed quark field solution would be, 
as usual, $\gamma_5 \, {\mathcal C} \, \psi^*(\tilde x)$. But in (\ref{fieq4}) 
the term containing $\bfpi$ has changed sign and we need to specify how the 
pion field transforms under TR in order to get the time-reversed quark 
solution. Under TR the equation for the pion field becomes 
\be
- \nabla^2 \bfpi - i \, N_c \, g \, \bar \psi'(x) \gamma_5 
\bftau^{\rm t} \psi'(x) + \frac{\partial U}{\partial \bfpi'} = 0 \,, 
\label{fieq5}
\ee
where primes denote time-reversed fields. 
By inspecting eqs.~(\ref{fieq4}) and (\ref{fieq5}) we recognize 
the existence of two possible realizations of TR operator.
The first one amounts to
keeping the standard TR for the quarks and 
reversing the sign of the $x$ and $z$ components of the pionic
field under TR. This realization of TR
is generally used in systems in which pions are emitted or absorbed
by a fermion in a perturbative way. The second one
consists in 
leaving the pion field unchanged and performing 
an isospin rotation on the quark field. Since
$\tau_2(-\bftau\,^{\rm t})\tau_2=\bftau$, the time-reversed
quark field is ($ab$ are isospin indices)
%
%
%
\be
T \psi_a(x) T^{-1}  = - (\tau_2)_{ab} 
\, \gamma_5 \, {\mathcal C} \, \psi_b^*(\widetilde{x}) \,, 
\label{TR34}
\ee
to be compared with the {\em standard} action of TR,
%
%
\be
T \psi_a(x) T^{-1} = - \gamma_5 \, {\mathcal C} 
\, \psi_a^*(\widetilde{x})\,.  
\label{TR2b}
\ee
The {\em non-standard} TR (\ref{TR34}) exhibits a unitary isospin 
rotation $\tau_2$, which is exactly of the form (\ref{TR6}) indicated 
by Weinberg, with $\phi = - \pi$. Thus the system of quarks and chiral 
fields governed by (\ref{TR7}) exemplifies Weinberg's unconventional  
representation of TR. 
The existence of various possibe definitions of time reversal
operator is due to the degeneracy of the ground state of our model.
It is also important to notice that none of the degenerate ground states
is a flavor eigenstate \cite{fiolhais}. From this viewpoint,
the use of the non-standard TR operator is very natural.

Note that, if we generalize the lagrangian (\ref{TR7}) to $SU(3)$, it 
is straightforward to show that there is no unitary irreducible 
$3 \times 3$ matrix 
that gives the time-reversed solution. This agrees with Weinberg's 
conclusion that non-standard TR can mix at most two components of the 
particle multiplet. 


\vspace{0.5cm}

\noindent
{\bf 3.} In most practical calculations of nucleon structure, the 
so-called hedgehog form of the pion field, $\bfpi$ = $\hat{\bfr}\,\phi(r)$, 
is adopted. In this case one is able to compute explicitely
the ground-state quark configuration, which turns out to have 
the spin-isospin structure \cite{fiolhais}
\be
\vert h \rangle= \frac{1}{\sqrt{2}}\, 
\left ( \vert u - \rangle- \vert d + \rangle \right )\,. 
\label{TR12}
\ee
This is not a state of fixed isospin, but rather an eigenstate (with 
zero eigenvalue) of the so-called grand-spin $\bfG=\bfJ + \bfI$ (spin + 
isospin). It is immediate to check that the hedgehog solution (\ref{TR12}) 
is indeed invariant under the non-standard TR. 

It is important to notice that the nucleon state built up from the 
chiral lagrangian (\ref{TR7}) satisfies the usual TR properties. 
In the baryon rest frame, starting from the mean-field solution which is 
a superposition of nucleon and delta, we project out a state with definite 
spin and isospin by means of \cite{many}
\be
P_{J_3 I_3}^J\equiv (-1)^{J+I_3} \> \frac{2J+1}{8\pi^2}
\int \! d^3\Omega \>
{{\mathcal D}^{J\textstyle{*}}_{J_3,-I_3}}(\Omega ) \, R(\Omega )\,,
\label{TR13}
\ee
where ${\mathcal D}^J_{J_3,- I_3}(\Omega )$ is the 
familiar Wigner function and $R(\Omega )$ is the rotation operator.
Due to the symmetry of the hedgehog (which has grand-spin zero), 
the rotation can be performed either on spin or on isospin. 
If we choose to perform a spin rotation, a spin-isospin baryon eigenstate 
({\it e.g.}, a nucleon) is obtained as
\be
\vert J,J_3,  I_3\rangle=P_{J_3 I_3}^J \vert H \rangle,
\label{TR14}
\ee
where $\vert H \rangle $ is the mean-field solution, made of three
quarks in the configuration (\ref{TR12}) surrounded by a coherent state 
of pions. 
We now want to check that the TR operator for the baryon state defined 
by (\ref{TR14}) is the usual one, that is $T = K \, \Theta(\sigma)$,  
where $K$ is the operator of complex conjugation and 
$\Theta (\sigma) = - i \sigma_2$. Under $T$, the projector transforms as
\bq
T P^J_{J_3 I_3} T^{-1} &=& (-1)^{J_3- I_3}P^J_{-J_3,-I_3}
\nonumber\\
&=&(-1)^{J-J_3} P^J_{-J_3, I_3} (- i \tau_2)\, .
\label{TR16}
\eq
Since the hedgehog state is invariant under the combined action of
$\Theta(\sigma) \, \Theta(\tau)$, applying (\ref{TR16})
to (\ref{TR14}) gives
\be
T P^J_{J_3 I_3}| H \rangle=
(-1)^{J-J_3} P^J_{- J_3, I_3}| H \rangle\, ,
\label{TR17}
\ee
which is the standard time-reversal transformation, as anticipated.

\vspace{0.5cm}

\noindent 
{\bf 4.}  Let us now explore the implications of the non-standard TR 
on the spin structure of hadrons. Although, as we have just seen,
the TR operator defined in (\ref{TR34}) does not affect the time-reversal 
properties of the nucleon as a whole, it does have important consequences 
on the internal quark dynamics. In particular, we shall see that the 
so-called ``$T$-odd''quark distribution functions, introduced 
by some authors \cite{sivers,mauro,boer1,boglione} to account 
for the single-spin asymmetries experimentally observed in transversely 
polarised pion hadroproduction \cite{adams}, are actually allowed by 
TR invariance, if the TR operator acting on the quark fields is the one 
given in (\ref{TR34}).

When used to constrain the general form of the quark-quark correlation 
matrix $\Phi$ in nucleons, which incorporates all quark distribution 
functions, TR invariance, implemented in the standard way via the operator 
(\ref{TR1b}), forbids correlations of the form 
\bq
& & \varepsilon^{\mu \nu \rho \sigma} \, P_{\nu} \, k_{\perp \rho} \,
S_{\perp \sigma}\,, 
\label{TR3} \\
& & \varepsilon^{\mu \nu \rho \sigma} \, P_{\rho} \, k_{\perp \sigma}\,, 
\label{TR4}
\eq
where $P, S$ are the proton's four-momentum and spin, respectively 
($\perp$ denoting the transverse components of $S$), and $\bfk_{\perp}$ 
is the transverse momentum of quarks. The leading-twist $T$-odd quark 
distributions arising from the terms in $\Phi$ of the form (\ref{TR3}) 
and (\ref{TR4}) are called $f_{1T}^\perp(x, \bfk_{\perp}^2)$ and 
$h_1^\perp(x, \bfk_{\perp}^2)$, respectively. The former is related to 
the number density of unpolarised quarks in a transversely polarised 
nucleon; the latter measures the transverse polarisation of quarks in 
an unpolarised hadron. If we call ${\mathcal P}_{q/p} (x, \bfk_{\perp})$ 
the probability to find a quark with momentum fraction $x$ and 
transverse momentum $\bfk_{\perp}$ in the proton, we have \cite{report}
\bq
& &  {\mathcal P}_{q/p^{\uparrow}} (x, \bfk_\perp) -
  {\mathcal P}_{q/p^{\uparrow}} (x, - \bfk_\perp)
  \nonumber \\   
& & \hspace{0.5cm} =
 -2 \, \frac{| \bfk_\perp |}{M} \, \sin(\phi_k - \phi_S)\,
  f_{1T}^\perp(x, \bfk_\perp^2) \nonumber \\
& & \hspace{0.5cm} \equiv 
 \sin(\phi_k - \phi_S) \, 
\Delta_0^T f(x, \bfk_{\perp}^2)
\,,
\label{TR30} 
\eq
\bq
& &  {\mathcal P}_{q^{\uparrow}/p}   (x, \bfk_\perp) -
  {\mathcal P}_{q^{\downarrow}/p} (x, \bfk_\perp) 
\nonumber \\
& & \hspace{0.5cm} =   
- \frac{| \bfk_\perp |}{M} \, \sin(\phi_k - \phi_S)
  \, h_1^\perp(x, \bfk_\perp^2) \nonumber \\
& & \hspace{0.5cm} \equiv 
 \sin(\phi_k - \phi_S)\, 
\Delta_T^0 f(x, \bfk_{\perp}^2)
\,,
\label{TR31}
\eq
where the arrows denote transverse polarisation states, and $\phi_k$ 
and $\phi_S$ are the azimuthal angles of $\bfk_{\perp}$ and 
$\bfS_{\perp}$, respectively. The field-theoretical definitions of 
$\Delta_0^T f$ and $\Delta_T^0 f$ are ($\vert P, \pm \rangle$ are the 
momentum-helicity eigenstates of the proton and $a$ is the flavor index)
\bq
\Delta_0^T f
(x, \bfk_{\perp}^2) &=& 
{\rm Im} \; \int \frac{{\rm d}y^- {\rm d}^2 \bfy_\perp}{2(2 \pi)^3}
\, {\rm e}^{-i x P^+ y^- + i \bfk_\perp\cdot \bfy_\perp} 
\nonumber \\
& & \times \langle P, - \vert \overline{\psi}_a(0,y^-, y_\perp)
{\gamma^+}\psi_a(0)\vert P, + \rangle \,, 
\label{TR32} 
\eq
\bq
\Delta_T^0 f(x, \bfk_{\perp}^2) &=& 
 \int \frac{{\rm d}y^- {\rm d}^2 \bfy_\perp}{ 2(2 \pi)^3}
\, {\rm e}^{-i x P^+ y^- + i \bfk_\perp\cdot \bfy_\perp} 
\nonumber \\
& & \times \langle P \vert \overline{\psi}_a(0,y^-,y_\perp)
{i \sigma^{2 +} \gamma_5 }\psi_a(0)\vert P \rangle \,.
\label{TR33}
\eq
%

The distribution $\Delta_0^T f$ was first introduced by Sivers \cite{sivers}   
and its phenomenological implications were investigated in 
\cite{mauro,boer1,boglione}; in some of these papers $\Delta_0^T f$ is 
denoted by $\Delta^Nf_{q/\pup}$. The distribution $\Delta_T^0 f$ was 
studied by Boer and Mulders \cite{boer1,boer2}.
Using the standard action of TR on quark fields, eq.~(\ref{TR2b}),  
it is easy to show that the matrix elements in (\ref{TR32}, \ref{TR33}), 
and the corresponding distributions, change sign under $T$, 
\be
T : \; \Delta_0^T f_q \to - \Delta_0^T f_q
 \,. 
\label{TR33b}
\ee
Therefore, invariance under standard TR implies that these distributions  
must vanish, as -- in a first paper -- pointed out by Collins \cite{collins}.

This conclusion would be inescapable if the quark fields were {\em free}. 
However, they are {\em interacting} fields and this renders the 
implementation of TR invariance rather subtle. 
If quark interactions are modelled by a chiral lagrangian like 
(\ref{TR7}), the TR operator acts as in (\ref{TR34}), and consequently  
TR transforms the $u$ quark distribution into minus the $d$ quark 
distribution, and viceversa: 
\be
T : \; \Delta_0^T f_u \to - \Delta_0^T f_d
 \,. 
\label{TR35}
\ee
A similar relation holds for $\Delta_T^0 f$. Therefore, what TR 
invariance entails is not the vanishing of the $u$ and $d$ distributions 
separately, but of their isoscalar combination
\be
\Delta_0^T f_{u} + \Delta_0^T f_{d} = 0 \,. 
\label{TR36} 
\ee

A caveat is in order. We cannot take the relation (\ref{TR36}) too 
literally. In fact, it arises  from the assumptions we made about the 
quark dynamics (in particular, the time-independence of the chiral 
fields). Hence, we should not expect (\ref{TR36}) to hold in general. 
What we do expect, however, is that non-perturbative quark dynamics
realizes TR in a non-standard way, so that our conclusion that the 
distributions $\Delta_0^T f$ and $\Delta_T^0 f$ are {\em not} 
necessarily forbidden by TR invariance should be general and quite firm. 
 
The same conclusion has been recently reached by Collins, who has 
reconsidered his proof of the vanishing of $\Delta_0^T f$  based on
(\ref{TR32}) and (\ref{TR33b}). The correct field-theoretical expressions 
of quark distributions contain link operators ({\it i.e.}, path-ordered 
exponentials of the gluon field, the so-called Wilson lines), 
which are usually omitted (being unity in the axial gauge), 
but have a non-trivial time-reversal behavior. It turns out that, 
under $T$, a future-pointing Wilson line is transformed into a 
past-pointing Wilson line, and hence the distribution $\Delta_0^T f$ 
does {\em not} simply change sign for time reversal. The conclusion 
(\ref{TR33b}) is therefore wrong and time reversal invariance, rather 
than constraining $\Delta_0^T f$ to zero, gives a relation between 
asymmetries probed in different processes \cite{collins2}.  

We have already mentioned that the $T$-odd  distribution functions have 
interesting phenomenological consequences. In particular, the non vanishing 
of $\Delta_0^T f$ could explain \cite{mauro} the azimuthal asymmetries 
in pion hadro-production observed by the E704 experiment \cite{adams}. 
In order to justify the existence of $T$-odd distribution functions, 
their proponents \cite{mauro,boer1,boglione} invoke initial-state effects 
which would produce non-trivial relative phases between the colliding 
hadrons, thus preventing implementation of the ``na{\"\i}ve'' 
time-reversal invariance via (\ref{TR1}). The situation would be specular 
to that occurring in the fragmentation process of single-inclusive 
lepto-production, where $T$-odd fragmentation functions are made possible 
by non trivial {\em final-state interactions}. It is quite difficult, 
however, to figure out a physical mechanism giving rise to 
{\em initial-state interactions}, but still preserving the QCD factorization. 
A corollary of this line of reasoning is that $T$-odd distributions 
should be observable only in reactions involving two initial hadrons 
({\it i.e.}, in Drell-Yan processes, hadron production in proton-proton 
collisions, etc.).  What we have shown here is that $\Delta_0^T f$ and 
$\Delta_T^0 f$ are {\em not} $T$-odd, if TR is implemented via (\ref{TR34}).
If our idea is correct, these distributions could be probed not only 
in pion {\em hadro}-production, but also in pion {\em lepto}-production.  
Focusing on $\Delta_0^T f$, we get a leading-twist contribution to the 
cross section for the semi-inclusive process $\ell p^{\uparrow} 
\to \ell \pi X$, which reads (for a review of transversely polarised 
leptoproduction of hadrons see \cite{report}) 
\bq
& & \frac{{\rm d} \sigma}{{\rm d}x \, {\rm d} y \, {\rm d}z \, 
{\rm d}^2 \bfP_{\perp} \, {\rm d} \phi} 
= \frac{2 \, \pi \, \alpha_{\rm em}^2}{Q^2} 
\, \frac{1 + (1 - y)^2}{y} \nonumber \\
&& \hspace{0.5cm} \times \,  \sum_{q, \bar q}
e_q^2 \, \Delta_0^T \, f_{q}(x, \bfP_{\perp}^2) \,  
D_{q}^{\pi}(z) \, \sin \phi \,,  
\label{TR38} 
\eq
where $D_{q}^{\pi}(z)$ is the fragmentation function of quarks into 
pions, and $\phi$ is the angle between the direction of the transverse 
spin of the target and the momentum $\bfP_{\!\perp}$ of the produced pion
(in the $\gamma^*$-$\,p$ c.m. frame). Some evidence of non-zero 
single-spin asymmetries in lepto-production has been reported by the 
SMC \cite{smc} and the HERMES Collaboration \cite{hermes}. 

\vspace{0.5cm}
\noindent
{\bf 5.} We have shown that a system of interacting quark and chiral 
fields provides a physical realization of a non-standard TR that mixes 
the components of an isospin multiplet.  As a consequence, the $T$-odd 
distribution functions of the nucleon are not obliged to vanish
and may generate single-spin asymmetries in lepto-production. Recently, 
Brodsky, Hwang and Schmidt \cite{brodsky} have proven that such 
asymmetries may arise at leading twist due to gluon exchange between 
the outgoing quark and the target spectator system. This contribution 
is not power-law suppressed in $Q^2$, and behaves as $1/\bfk_{\perp}^2$.  
Collins \cite{collins2} has pointed out that the mechanism 
proposed by Brodsky {\it et al.} is compatible with QCD factorization and 
can be ascribed to a transverse-spin asymmetry in the $\bfk_{\perp}$ 
distribution of quarks, that is to a $\Delta_0^T f$ function. 
Although Collins argument for the non vanishing of $\Delta_0^T f$ 
is quite different from the one presented in this paper, it has has an 
important point in common with it: both mechanisms 
show that quark interactions (in the form of gluon exponentials in one 
case, of chiral fields in the other case) deeply affect the time-reversal 
behavior of distribution functions. 
It is also important to remark that distribution functions are usually 
computed using effective lagrangians,
like the chiral model one, which are appropriate at a low momentum scale.
The argument presented in \cite{collins2} on the non-vanishing
of $T$-odd distribution functions is based on QCD and involves the
dynamics of the gluonic degrees of freedom. On the other hand,
in model lagrangians, the effective degrees of freedom are different
from the ones of QCD and, for instance, chiral degrees of freedom
can take the place of gluonic ones. It is therefore important to
show how the argument presented in \cite{collins} can be bypassed
without making an explicit use of gluon dynamics. The result presented
here goes along these lines, offering a realization of the idea
discussed in \cite{collins2} in terms of chiral degrees of freedom.

\end{document}